\documentclass{article}

\usepackage{amssymb,amsfonts,amsmath,stmaryrd}
\usepackage{cite,enumerate,float,indentfirst}
\usepackage{color}
\usepackage{tikz}
\usetikzlibrary{arrows,snakes,backgrounds}

\def\be{\begin{eqnarray}}
\def\ee{\end{eqnarray}}
\def\nn{\nonumber}

\def\p{\partial}
\def\tr{{\rm tr}\,}
\def\Tr{{\rm Tr}\,}

\def\S{{\rm Schur}}

\definecolor{red}{rgb}{1,0,0}
\definecolor{orange}{rgb}{1,0.5,0}
\definecolor{violet}{rgb}{0.7,0,1}

\hoffset= - 1.0in         % switch off for draft style
\begin{document}

\begin{center}
\begin{small}
\hfill MIPT/TH-16/22\\
\hfill FIAN/TD-10/22\\
\hfill ITEP/TH-19/22\\
\hfill IITP/TH-18/22\\

\end{small}
\end{center}

\vspace{.5cm}

\begin{center}
\begin{Large}\fontfamily{cmss}
\fontsize{15pt}{27pt}
\selectfont
	\textbf{Superintegrability as the hidden origin of Nekrasov calculus}
	\end{Large}
	
\bigskip \bigskip

\begin{large}
A. Mironov$^{b,c,d}$\,\footnote{mironov@lpi.ru; mironov@itep.ru},
A. Morozov$^{a,c,d}$\,\footnote{morozov@itep.ru}
\end{large}

\bigskip

\begin{small}
$^a$ {\it MIPT, Dolgoprudny, 141701, Russia}\\
$^b$ {\it Lebedev Physics Institute, Moscow 119991, Russia}\\
$^c$ {\it Institute for Information Transmission Problems, Moscow 127994, Russia}\\
$^d$ {\it NRC ``Kurchatov Institute'' - ITEP, Moscow 117218, Russia}\\
\end{small}
 \end{center}

\bigskip

\begin{abstract}
Once famous and a little mysterious, AGT relations between Nekrasov functions
and conformal blocks are now understood as the
Hubbard-Stratanovich duality in the Dijkgraaf-Vafa (DV) phase
of a peculiar Dotsenko-Fateev multi-logarithmic matrix model.
However, it largely remains a collection of somewhat technical tricks,
lacking a clear and generalizable conceptual interpretation.
Our new claim is that the Nekrasov functions emerge in matrix models
as a straightforward implication of superintegrability,
factorization of peculiar matrix model averages.
Recently, we demonstrated that, in the Gaussian Hermitian model,
the factorization property can be extended
from averages of single characters to their bilinear combinations.
In this paper, we claim that this is true also for
multi-logarithmic matrix models, where factorized
are just the point-split products of two characters.
It is this {\it enhanced superintegrability} that is responsible for
existence of the Nekrasov functions and the AGT relations.
This property can be generalized both to multi-matrix models,
thus leading to AGT relations for multi-point conformal blocks,
and to DV phases of other non-Gaussian models.
\end{abstract}

\bigskip

\section{Introduction}

Once upon a time, in \cite{MMShproof},
we explained that the deep {\it raison d'etre}
for Nekrasov functions \cite{Nek}
and AGT relations \cite{AGT} to exist
are the Kadell formulas \cite{Kan,Kad,Kad2}
for the Selberg integrals \cite{Selb},
i.e. factorization of {\it pair} averages of Schur functions
with the double-logarithmic Selberg measure (see sec.\ref{fdc}),
\be
\Big<S_{R'}\{p_k+v\}\cdot S_{R''}\{p_k\}\Big> \ \equiv
\prod_{i=1}^{N} \int_0^1 dx_i   x_i^u (1-x_i)^v \Delta^2(x)
S_{R'}\{p_k+v\} S_{R''}\{p_k\}
\label{aveJackpair}
\ee
with $p_k=\sum_{i=1}^N x_i^N$.
The Nekrasov function is nothing but
\be
{\cal N}_{R'R''} = \ \Big<S_{R'}\{p_k+v\}\cdot S_{R''}\{p_k\}\Big>_{_+} \Big<S_{R'}\{p_k+v\}\cdot S_{R''}\{p_k\}\Big>_{_-}
\ee
with two different sets of the $u,v$-variables denoted by $+$ and $-$,
or with two different integration contours, associated with DV phase of the three-logarithm matrix model \cite{MMSh1}.

\bigskip

Nowadays we identified the property that an average of Schur-like functions (characters)
is factorized into a product of similar functions as {\it superintegrability},
pertinent to specific class of matrix and string models,
see \cite{MMSI} for a recent review and an ample list of references.
This factorization strengthens the usual integrability of matrix and eigenvalue models \cite{UFN3},
and, generically, of any non-perturbative functional integrals \cite{UFN2},
and resembles the closeness of orbits for quadratic and Coulomb potentials,
which are peculiar {\it superintegrable} cases of the generically integrable motion
in any central potential \cite{SI}.

\bigskip

It is a natural question to ask what is the connection of these two properties,
and if the Selberg measure possesses some {\it enhanced superintegrability},
which applies to {\it pair} correlators of characters
and is responsible for the very existence of the (still partly mysterious) Nekrasov calculus.
This latter deals with the Nekrasov functions, which are nicely factorized quantities
providing a new class of objects in representation theory,
they are still awaiting a clear non-technical definition.

\bigskip

The usual problem with pair correlators is that the product of characters
is expanded in characters, but gives rise to a linear combination with sophisticated
Littlewood-Richardson coefficients, not to a single term \cite{Mac}.
Thus, within the superintegrable context,
the average is also a linear combination of factorized quantities,
which is not expected to, and generically does {\it not}, factorize.
The Kadell formulas, however, demonstrate that, for a particular
double-logarithmic Dotsenko-Fateev
matrix model \cite{MMSh1,MMMSh}, they {\it do},
at expense of a simple point-splitting by $v$ in (\ref{aveJackpair}).
This is a drastic simplification as compared to the recently analyzed
Gaussian Hermitian model where certain bilinear combinations of characters
were also shown to factorize \cite{MMdsi},
but the simple point-splitting interpretation is not available.

\bigskip

In fact, after the $\beta$-deformation
(and further after the full-fledged $q,t$-deformation)
factorized are both pair correlators of the Jack polynomials and also the averages of {\it generalized} Jack polynomial,
and these latter are responsible for emergence of the standard Nekrasov functions.

\bigskip

In this paper, we propose to call Nekrasov function a product of two different Dijkgraaf-Vafa (DV) blocks, which are defined to be
bilinear combination of characters (symmetric functions) factorized with a proper averaging. This allows us to associate a matrix (eigenvalue) model with a given set of Nekrasov functions. The DV blocks giving rise to the standard Nekrasov functions are obtained as averages of point-split bilinear product of two Schur functions in the non-deformed case, and as averages of  generalized Jack polynomials, which are bilinear combinations of the Jack polynomials in the deformed one.We also consider other factorized averages of simple bilinear combinations of characters, they can be also associated with Nekrasov functions, and we explain what matrix (eigenvalue) models they give rise to.

This means that the notion of Nekrasov function turns out to be intimately related to factorization properties of averaging, and this is exactly what is known under the name of superintegrability of matrix models. Moreover, the Nekrasov functions are related this way with a more subtle avatar of superintegrability, with the factorization property of bilinear correlators of characters. This latter is basically a new issue in superintegrable theories.

The paper is organized in the following way. In section 2, we consider the matrix model with potential that is a sum of three logarithm terms. This model is a Dijkgraaf-Vafa type model, and it leads to the standard Nekrasov functions, which we demonstrate in section 3. In section 4, we discuss the $\beta$-deformation of this matrix model, and realize that the corresponding Nekrasov functions are given by averages of the generalized Jack polynomials. At last, in section 5, we discuss eigenvalue models that originate from the Nekrasov functions associated with averages of point-split products of Jack polynomials, and with the matrix model emerging from the Nekrasov functions associated with factorized bilinear correlator in the Gaussian matrix model. Section 6 contains some concluding remarks.

Altogether, we reproduce results about logarithmic models, which are partly known from \cite{Kan,Kad,Kad2,MMShproof,AFLT,MS,MMZ}
but now we rewrite and interpret them differently. This new interpretation provides us with a unified look at Nekrasov functions:
as factorized bilinear character correlators.
This gives rise to various new generalizations mostly discussed in sec.\ref{newm}, and, no less important,
to links with non-logarithmic matrix models in sec.\ref{Gauss},
where factorization is far less obvious and technically more involved.

To put differently, {\bf (i)} we start from bilinears in characters that have factorized averages; {\bf (ii)} this allows us to construct DV blocks; {\bf (iii)} products of two DV blocks define Nekrasov functions; {\bf (iv)} this gives rise to a matrix or eigenvalue model. The construction immediately extends to multi-linear products of DV blocks. This is one of the main new results of this paper. 

Another new result is expressed in formulas (\ref{dcJ}), (\ref{dGJ+J}) and (\ref{fK}). It supports our conjecture that the factorized averages of bilinears of characters are expressed through the same characters.

\paragraph{Notation.} Throughout the paper, we use the notation $S_R\{P_k\}$ for the Schur functions \cite{Mac}. It is a graded symmetric polynomial  of variables $\xi_i$, or of the power sums $P_k=\sum_i\xi_i^k$. The Schur functions are labelled by partitions (Young diagrams) $R:\ R_1\ge R_2\ge\ldots\ge R_l$ with $l_R$ parts (lines of the diagram), $|R|:=\sum_iR_i$. We also use the notation $S_R\{x\}:=S_R\{P_k=x\}$ and $d_R:=S_R\{\delta_{k,1}\}$. Similarly, we use the notation $J_R\{P_k\}$ for the Jack polynomials \cite{Mac}. We normalize the Jack polynomials as in \cite{MMShrev}. We denote skew Schur function $S_{R/Q}\{P_k\}$ and skew Jack polynomials $J_{R/Q}\{P_k\}$.

\section{DV phases of the three-logarithm model}

\subsection{DV phases of matrix models}

Consider the Hermitian one-matrix model with a potential $V(H)$,
\be\label{Zmm}
Z_N\{V;P_k\}\sim\int dH\exp\left(-\Tr V(H)+\sum_k{P_k\over k}\Tr H^k\right)
\ee
where the integration goes over $N\times N$ matrix $H$, and $dH$ is the Haar measure on Hermitian matrices. The integral is understood as a power series in variables $P_k$'s, hence it is given by moments of  invariant polynomials of $H$ with measure $\exp\left(-\Tr V(H)\right)$. As soon as the integrand of (\ref{Zmm}) is an invariant function, one can integrate over the angular variables to obtain \cite{Mehta}
\be\label{Zev}
Z_N\{V;P_k\}\sim \prod_{i=1}^N\int dh_i\exp\left(-V(h_i)+\sum_k{P_k\over k}h_i^k\right)\Delta^2(h)
\ee
where $h_i$'s are eigenvalues of matrix $H$, and $\Delta (h)$ is the Vandermonde determinant. The integration contour is chosen in such a way that the integral converges.

The Dijkgraaf-Vafa phase \cite{DV,DVmore,DVmore2} emerge when the potential $V(h)$
possesses several different extrema at points $h = \alpha_r$, $r=1,\ldots,s$. Then \cite{KMT,Mir,MMZ12}, there are $s$ different independent integration contours $K_r$ such that the integral (\ref{Zev}) converges. These contours definitely form a linear space, since one may choose a linear combination of them. In fact, one may define the DV partition function so that $N_r$ out of $N$ eigenvalues $h_i$ are integrated along $K_r$.
These $N_r$ serve as the $s$ additional moduli, if the answer is analytically
continued from the integer values of $N_r$ to arbitrary ones.
Thus, one can define the Dijkgraaf-Vafa partition function
$Z_{N_1,\ldots,N_s}\{t_k\}$ as a matrix (or, better to say, eigenvalue)
model with $s$ different integration contours so that $\sum_{r=1}^sN_r=N$ integration variables $h_i$ are parted into $s$ subsets which we denote by the superscript of $h$: the eigenvalues with $i=1,\ldots,N_1$ are integrated over $K_1$, we denote them $h_i^{(1)}$, $i=1,\ldots,N_1$; those with $i=N_1+1,\ldots,N_1+N_2$ are integrated over $K_2$, we denote them $h_i^{(2)}$, $i=1,\ldots,N_2$, etc.:
\be
Z_{N_1,\ldots,N_s}\{P_k\} \sim \prod_{r=1}^s\prod_{i=1}^{N_r}\int_{K_r}dh_i^{(r)} \exp\left(-V(h^{(r)}_i)+\sum_k{P_k\over k}\Big(h_i^{(r)}\Big)^k\right) \Delta^2(h)
\label{ZDV}
\ee

In this paper, we mostly concentrate on the case of potential with two minima, and consider partition functions at $P_k=0$ so that they are just functions of parameters of the potential $V(h)$ and of $N_1$, $N_2$, and apply the technique of superintegrability. The case of non-zero $P_k$'s requires further refinement of superintegrability approach (see, e.g., \cite{MMZ3}), and will be discussed elsewhere.

\subsection{Three-logarithm model}

In order to demonstrate the phenomenon that we discuss throughout the paper, we consider the matrix model with the potential that is a sum of three logarithms, since this model admits very explicit calculations. This model is inherited from the Dotsenko-Fateev \cite{DF} representation \cite{AGTmamo,MMSh1,MMMSh} of the conformal blocks in $2d$ conformal field theory, and is given by the matrix integral over $N\times N$ Hermitian matrix $H$
\be
Z_N(\{\alpha_a\};\{w_a\})\sim\int dH\exp\left(\Tr \Big(\sum_{i=a}^3\alpha_a\log (H-w_a)\right)
\ee
Here $dH$ is the Haar measure on the Hermitian matrices, and $\{\alpha_a,w_a\}$, $a=1,2,3$ are parameters. By a rescaling and a constant shift of $H$, one can achieve $w_1=0$, $w_2=1$, and we will denote $w_3:=w$.

After integrating out the angular variables, one can reduce this matrix integral to the integral over eigenvalues of $H$
\be\label{3l}
Z_N(\{\alpha_a\},w)\sim\prod_{i=1}^N \int dh_i h_i^{\alpha_1} (1-h_i)^{\alpha_2}(w-h_i)^{\alpha_3} \Delta^2(h)
\ee
Partition function (\ref{3l}) is associated with the potential
\be
V(h)=\alpha_1\log h+\alpha_2\log (1-h)+\alpha_3\log(w-h)
\ee
This potential has two minima, and, in accordance with the general rule above, there are two independent contours. We choose them to be $C_1:=[0,w]$ and $C_2:=[1,\infty)$. Thus, the $N$ integrations in (\ref{3l}) are parted between these two contours.
Let us assume $N_+$ variables $h_i$, $i=1,\ldots,N_+$ runs over contour $C_1$, and the remaining $N_-=N-N_+$, over contour $C_2$, and
change the variables: $h_i=wx_i$, $i=1,\ldots N_1$, $h_{N_1+j}=y_j^{-1}$. Then, integral (\ref{3l}) becomes
\be\label{2Sel}
Z_{N_1,N_2}(\{\alpha_a\};w)={\cal C}\cdot\prod_{i=1}^{N_+} \int dx_i x_i^{\alpha_1} (1-x_i)^{\alpha_3}(1-wx_i)^{\alpha_2} \Delta^2(x)
\times \nn\\
\times \prod_{j=1}^{N_-} \int dy_j y_j^{-\alpha_1-\alpha_2-\alpha_3-2N} (1-y_j)^{\alpha_2}(1-wy_j)^{\alpha_3} \Delta^2(y)
\times \prod_{i,j}(1-wx_iy_j)^2
\ee
and we fix the normalization ${\cal C}$ so that $Z_{N_1,N_2}(\{\alpha_a\};0)=1$. Later on, we use the traditional notation $u_+=\alpha_1$, $v_+=\alpha_3$, $u_-=-\alpha_1-\alpha_2-\alpha_3-2N$, $v_-=\alpha_2$ so that there is a constraint for these four variables
\be\label{cnst}
2N+u_++v_++u_-+v_-=0
\ee

\subsection{Selberg model as an elementary building block}

The representation (\ref{2Sel}) of the three-logarithm model implies that it can be constructed from the two building elementary blocks, which we will call the DV building blocks, each of them being associated with one concrete contour. In other words, we represent (\ref{2Sel}) as an average in two {\it distinct} matrix models of the usual (non DV) type:
\be\label{CB}
Z_{N_1,N_2}(u_\pm,v_\pm;w)&:=&\left< \ \left< \ \ \prod\limits_{i = 1}^{N_+} (1 - w x_i)^{v_-} \prod\limits_{j = 1}^{N_-} (1 - w y_j)^{v_+} \prod\limits_{i = 1}^{N_+} \prod\limits_{j = 1}^{N_-} (1 - w x_i y_j)^{2} \ \right>_+ \ \right>_-
\ee
where the two matrix model averages are given by Selberg-type integrals
\be\label{Sav}
\Big< f \Big>_+ \ = \
\int\limits_{0}^{1} dx_1 \ldots \int\limits_{0}^{1} dx_{N_+} \prod\limits_{i<j} (x_i - x_j)^{2} \prod\limits_{i} x_i^{u_+} (x_i - 1)^{v_+} \ f\big(x_1, \ldots, x_{N_+}\big)\nn\\
\Big< f \Big>_- \ = \
\int\limits_{0}^{1} dy_1 \ldots \int\limits_{0}^{1} dy_{N_-} \prod\limits_{i<j} (y_i - y_j)^{2} \prod\limits_{i} y_i^{u_-} (y_i - 1)^{v_-} \ f\big(y_1, \ldots, y_{N_-}\big)
\ee
and the averages are as usual normalized in such a way that $\Big< 1 \Big>_+ = \Big< 1 \Big>_- = 1$.

In order to calculate this average, we use the transformation \cite{Ito}
\be\label{exp}
\prod\limits_{i = 1}^{N_+} (1 - w x_i)^{v_-} \prod\limits_{j = 1}^{N_-} (1 - w y_j)^{v_+} \prod\limits_{i = 1}^{N_+} \prod\limits_{j = 1}^{N_-} (1 - w x_i y_j)^{2} = \emph{}\nn\\
\emph{} = \exp\left( - \sum\limits_{k=1}^{\infty} \dfrac{w^k}{k} {\widetilde p}_k( p_k +\beta^{-1} v_+) \right) \exp\left( - \beta \sum\limits_{k=1}^{\infty} \dfrac{w^k}{k} p_k( {\widetilde p}_k + \beta^{-1}v_-) \right)
\ee
where $p_k = \sum_i x_i^k$ and ${\widetilde p}_k = \sum_i y_i^k$.

Using the Cauchy identity, the r.h.s. of (\ref{exp}) can be expanded to a sum bilinear in the Schur functions:
\be\label{Ce}
 \exp\left( - \sum\limits_{k=1}^{\infty} \dfrac{w^k}{k} {\widetilde p}_k( p_k + v_+) \right) \exp\left( - \sum\limits_{k=1}^{\infty} \dfrac{w^k}{k} p_k( {\widetilde p}_k + v_-) \right) =\nn\\
\emph{} = \sum\limits_{R',R''} w^{|R'|+|R''|} S_{R'}\{ - p_k - v_+ \} S_{R''}\{p_k\} S_{R'}\{ {\widetilde p}_k \} S_{R''}\{ -  {\widetilde p}_k - v_- \}
\ee
Now the averages in (\ref{CB}) split into products of two Selberg averages \cite{MMShproof}:
\be\label{Zsum}
Z_{N_1,N_2}(u_\pm,v_\pm;w)=\sum\limits_{R',R''} w^{|R'|+|R''|} \underbrace{\Big<S_{R'}\{ - p_k - v_+ \} S_{R''}\{p_k\}\Big>_+
\Big<S_{R'}\{ {\widetilde p}_k \} S_{R''}\{ -  {\widetilde p}_k - v_- \}\Big>_-}_{\rm Nekrasov\ functions}
\ee
These Selberg averages are factorized {\it due to the superintegrability}, and are called Nekrasov functions.

Hence, this illustrates our general approach to the DV type matrix model: one first has to part the partition function of this matrix model into DV building blocks and then, using superintegrability to evaluate contributions of factors mixing between the parts. These contributions are typically presented as series in (factorized) expressions, which are called Nekrasov functions.
These mixing terms can be dealt with in different ways, each of them leading to a different set of Nekrasov functions. The invariant notion is the DV matrix model, and, in this concrete case, the integral (\ref{CB}), which can be associated with the LMNS integral \cite{LMNS}, or with the 4-point conformal block \cite{AGT}, while the Nekrasov functions are associated with a very concrete expansion of these.

\section{Nekrasov functions and three-logarithm model}
\subsection{Factorization of double correlators}

In this section, we demonstrate that superintegrability implies a factorization of proper double correlators of the Schur functions, which allows one to rewrite (\ref{Zsum}) as an explicit series with coefficients being the standard Nekrasov functions.

First of all, note that superintegrability implies that the single average is factorized (as usual, our normalization of measure is such that $\Big<1\Big>=1$) \cite{Kan,Kad}:
\be\label{1cor}
\Big<S_R\{p_k\}\Big> :=\prod_{i=1}^N \int_0^1 dx_i x_i^u (1-x_i)^v \Delta^2(x)
S_{R}\Big\{p_k=\sum_ix_i^k\Big\}= \frac{S_R\{N\}\cdot S_R\{N+u\}}{S_R\{2N+u+v\}}
\ee

\subsubsection{Factorization property of double correlators}

The product of two characters is a sum of characters with the coefficients
made from the Littlewood-Richardson coefficients and the skew-characters at the point-splitting
parameter  $\zeta$:
\be
\S_{R'}\{p\} \cdot \S_{R''}\{p+\zeta\}
=   \S_{R'}\{p\} \cdot \sum_Q  \S_{Q}\{p\} \cdot \S_{R''/Q}[\zeta]
= \nn \\
= \sum_R \left(\sum_Q N_{R'Q}^R\cdot  \S_{R''/Q}[w]\right)\cdot  \S_R\{p\}
\ee
where $N_{R'Q}^R$ are the Littlewood-Richardson coefficients. The question is what is so special about the coefficient in brackets,
and why it is so nicely adjusted
to the Selberg measure in (\ref{aveJackpair})?

Nothing of this kind happens in Gaussian Hermitian matrix model:
point splitting is not enough to make pair correlators factorized,
e.g.
\be
\Big<\S_{[2]}\{p\}\cdot\S_{[2]}\{p+\zeta\}\Big>_{GH}
= \Big<\S_{[4]} \Big>_{GH} + \Big<\S_{[31]} \Big>_{GH} + \Big<\S_{[22]} \Big>_{GH}
+ \nn \\
\underbrace{+ \zeta\cdot \Big<\S_{[3]} \Big>_{GH} + w\cdot \Big<\S_{[21]} \Big>_{GH}}_0
+ \frac{\zeta(\zeta+1)}{2}\cdot \Big<\S_{[2]} \Big>_{GH}
%= \nn \\
= \frac{N(N+1)(N^2+N+ \zeta^2+\zeta+4)}{4}
\ee
which does not  factorize  for any reasonable choice of $\zeta$,
while the same average in the Selberg case is much more complicated for generic $\zeta$,
but is nicely factorized for $\zeta=v$:
\be
\Big<\S_{[2]}\{p\}\cdot\S_{[2]}\{p+\zeta\}\Big>
= \Big<\S_{[4]} \Big> + \Big<\S_{[31]} \Big> + \Big<\S_{[22]} \Big>
+ \nn \\
+ \zeta\cdot \Big<\S_{[3]} \Big> + \zeta\cdot \Big<\S_{[21]} \Big>
+ \frac{\zeta(\zeta+1)}{2}\cdot \Big<\S_{[2]} \Big>
= \nn \\
= \frac{N(N+1)(N+u)(N+u+1)(N+v)(N+v+1)(N+u+v)(N+u+v+1)}{4(2N+u+v)^2(2N+u+v-1)(2N+u+v+3)}
\ee
A simpler version of the latter example is factorization for $\zeta=v$ of the average
\be
\Big<\S_{[1]}\{p\}\cdot\S_{[1]}\{p+\zeta\}\Big>
=\Big<\S_{[2]}\Big>+ \Big<\S_{[1,1]}\Big> + \zeta\cdot \Big<\S_{[1]}\Big>
= \nn \\
= \frac{N(N+u)(N+v)(N+u+v)}{(2N+u+v+1)(2N+u+v-1)}
\ee
where expression for generic $w$ is short enough to be presented, see (\ref{DF11}) below.
This example is not interesting in the Gaussian case:
the average is just $N$, because $\Big<\S_{[1]} \Big>_{GH}=0$,
as all Gaussian averages of the Schur functions for the diagrams of odd sizes.

In fact, there are bilinears in the Gaussian model, which are factorized, but they are
more complicated \cite{MMdsi} (see sec.\ref{Gauss}).
Thus bilinear factorization has a chance to be a generically present enhancement
and even corollary of superintegrability,
but only in the Selberg case it is described by a point-splitting.

\subsubsection{Factorized double correlators\label{fdc}}

Now let us concentrate on the double correlators of the Selberg averages. The simplest example is
\be
\prod_{i=1}^N \int_0^1 dx_i x_i^u (1-x_i)^v \Delta^2(x)
S_{[1]}\{p\} S_{[1]}\{p+\zeta\}={N(u+N)\Xi\over (u+v+2N-1)(u+v+2N)(u+v+2N+1)}\nn\\
\Xi:=\zeta(u+v)^2+N(u+v)(u+4\zeta)+(4N^2-1)\zeta+2N^3+3N^2u+(N^2+1)v
\label{DF11}
\ee
This expression is factorized at $\zeta=v$:
\be
\Xi=(v+N)(u+v+N)(u+v+2N)
\ee

The general formula at this value of $\zeta$ is still factorized \cite{Kad2,MMShproof}:
\be
<S_{R'}\{p_k+v\}\cdot S_{R''}\{p_k\}> \ \equiv
\prod_{i=1}^N \int_0^1 dx_i x_i^u (1-x_i)^v \Delta^2(x)
S_{R'}\{p_k+v\} S_{R''}\{p_k\} = \nn\\
= {\eta_{R'R''}\over \eta_{\emptyset \emptyset}}{S_{R'}\{v+N\}S_{R''}\{u+N\}\over S_{R'}\{N\}S_{R''}\{u+v+N\}}
\label{aveSpair}
\ee
where
\be
\eta_{R'R''}:={\prod_{i<j}^N(R'_i-i-R'_j+j)\cdot \prod_{i<j}^N(R''_i-i-R''_j+j)\over \prod_{i,j}^N(u+v+2N+1+R'_i-i+R''_j-j)}
\ee
Using the identity
\be
{\eta_{\emptyset \emptyset}\over \eta_{R'R''}}{S_{R'}\{x+N\}S_{R''}\{x+N\} S_{R'}\{N\}S_{R''}\{N\}}=
(-1)^{|R'|}G_{R''R'}(x+2N)G_{R'^\vee R''^\vee}(-x-2N)d_{R'}d_{R''}
\ee
with
\be
G_{R'R''}(x):=\prod_{(i,j)\in R'}(x+1+R'_i-i+R''_j-j)
\ee
one obtains that
\be\label{dcor}
<S_{R'}\{p_k+v\}\cdot S_{R''}\{p_k\}> \ ={(-1)^{|R'|}S_{R'}\{v+N\}S_{R''}\{u+N\} S_{R'}\{u+v+N\}S_{R''}\{N\}
\over G_{R''R'}(u+v+2N)G_{R'^\vee R''^\vee}(-u-v-2N)d_{R'}d_{R''}}
\ee
This explains how to deal with a particular case of $R'=\emptyset$, when the double correlator is reduced to a single one: using $G_{R\emptyset}(x)=S_R\{x\}d_R^{-1}$, one obtains from (\ref{dcor}) at $R'=\emptyset$ formula (\ref{1cor}).
Note also that $G_{RR^\vee}(0)=d_R^{-1}$, therefore one can rewrite (\ref{dcor}) in the form
\be\label{dcor2}
\boxed{
<S_{R'}\{p_k+v\}\cdot S_{R''}\{p_k\}> \ ={(-1)^{|R'|}S_{R'}\{v+N\}S_{R''}\{u+N\} S_{R'}\{u+v+N\}S_{R''}\{N\}
\over {\cal F}_{R'R''}(u+v+2N)d_{R'}^2d_{R''}^2}}\nn\\
\nn\\
{\cal F}_{R'R''}(x):=G_{R''R''^\vee}(0)G_{R''R'}(x)G_{R'^\vee R''^\vee}(-x)G_{R'R'^\vee}(0)
\ee
The denominator of (\ref{dcor2}) is proportional to $z_{vect}^{1/2}$, where $z_{vect}$ is the standard vector Nekrasov function,
\be
z_{vect}(R',R'',x)&=&{\cal F}_{R'R''}(x)^2
\ee
Note that there is a simple rule of transposition of the Young diagram for the Schur function, which allows one to change the sign of all time variables at once:
\be
S_{R^\vee}\{p_k\}=(-1)^{|R|}S_R\{-p_k\}
\ee
Hence, formula (\ref{dcor2}) becomes in this case
\be\label{dcor3}
\boxed{
<S_{R'}\{-p_k-v\}\cdot S_{R''}\{p_k\}> \ ={S_{R'}\{-v-N\}S_{R''}\{u+N\} S_{R'}\{-u-v-N\}S_{R''}\{N\}
\over {\cal F}_{R'^\vee R''}(u+v+2N)d_{R'}^2d_{R''}^2}}
\ee

\subsection{Nekrasov functions and AGT}

Now we use can immediately evaluate (\ref{Zsum}). Indeed, from (\ref{dcor3}) one obtains
\be\label{zi}
Z_{N_1,N_2}(u_\pm,v_\pm;w)=\sum\limits_{R',R''} w^{|R'|+|R''|}{S_{R'^\vee}\{v_++N_+\}S_{R''}\{u_++N_+\} S_{R'^\vee}\{u_++v_++N_+\}S_{R''}\{N_+\}
\over {\cal F}_{R'^\vee R''}(u_++v_++2N_+)d_{R'^\vee}^2d_{R''}^2}\times\nn\\ \times
{S_{R''^\vee}\{v_-+N_-\}S_{R'}\{u_-+N_-\} S_{R''^\vee}\{u_-+v_-+N_-\}S_{R'}\{N_-\}
\over {\cal F}_{R''^\vee R'}(u_-+v_-+2N_-)d_{R'}^2d_{R''^\vee}^2}
\ee
Using the notations
\be
2\mu_1:=v_+-u_+,\ \ \ 
2\mu_2:=v_++u_+,\ \ \ 
2\mu_3:=v_--u_-,\ \ \ 
2\mu_4:=v_-+u_-,\ \ \ 
\mu:=N_++\mu_2=-N_--\mu_4
\ee
and
\be
z_f(R,x):={S_R\{p_k=x\}\over d_R}
\ee
which describes contributions of the fundamental matter to the Nekrasov functions ($\mu_a$'s are proportional to masses of the fundamental hypermultiplets),
one rewrites (\ref{zi}) in the standard Nekrasov form \cite{MMShproof}
\be
Z_{N_1,N_2}(u_\pm,v_\pm;w)=\sum\limits_{R',R''} w^{|R'|+|R''|}{\prod_{a=1}^4z_f(R'^\vee,\mu_a+\mu)z_f(R''^\vee,\mu_a-\mu)\over
{\cal F}_{R'^\vee R''}(u_++v_++2N_+){\cal F}_{R''^\vee R'}(u_-+v_-+2N_-)}{d_{R'^\vee}^2d_{R''^\vee}^2\over
d_{R'}^2d_{R''}^2}=
\ee
\vspace{-0.3cm}
\be
=\sum\limits_{R',R''} w^{|R'|+|R''|}{\prod_{a=1}^4z_f(R'^\vee,\mu_a+\mu)z_f(R''^\vee,\mu_a-\mu)\over
{\cal F}_{R'^\vee R''}(u_++v_++2N_+)^2}=\sum\limits_{R',R''} w^{|R'|+|R''|}{\prod_{a=1}^4z_f(R',\mu_a+\mu)z_f(R'',\mu_a-\mu)\over
z_{vect}(R',R''^\vee,2\mu)}
\nn\ee
since $d_R=d_{R^\vee}$ and ${\cal F}_{R''^\vee R'}(u_-+v_-+2N_-)={\cal F}_{R''^\vee R'}(-u_+-v_+-2N_+)={\cal F}_{R'^\vee R''}(u_++v_++2N_+)$.

\section{$\beta$-deformation of three-logarithm model}

Now we will consider the $\beta$-ensemble deformation of the three-logarithm matrix model. This example is important as an illustration of an ambiguity of the Cauchy expansion when there are several natural bases of symmetric polynomials, and only one of them leads to factorized quantities (or to Nekrasov functions). The model is now given by the integral
\be\label{3lb}
Z_N^\beta(\{\alpha_a\},w)&\sim&\prod_{i=1}^N \int dh_i h_i^{\alpha_1} (1-h_i)^{\alpha_2}(w-h_i)^{\alpha_3} \Delta^{2\beta}(h)\nn\\
&:=&\left< \ \left< \ \ \prod\limits_{i = 1}^{N_+} (1 - w x_i)^{v_-} \prod\limits_{j = 1}^{N_-} (1 - w y_j)^{v_+} \prod\limits_{i = 1}^{N_+} \prod\limits_{j = 1}^{N_-} (1 - w x_i y_j)^{2\beta} \ \right>_+ \ \right>_-
\ee
with Selberg averages
\be\label{Savb}
\Big< f \Big>_+ \ = \
\int\limits_{0}^{1} dx_1 \ldots \int\limits_{0}^{1} dx_{N_+} \prod\limits_{i<j} (x_i - x_j)^{2\beta} \prod\limits_{i} x_i^{u_-} (x_i - 1)^{v_-} \ f\big(x_1, \ldots, x_{N_+}\big)\nn\\
\Big< f \Big>_- \ = \
\int\limits_{0}^{1} dy_1 \ldots \int\limits_{0}^{1} dy_{N_-} \prod\limits_{i<j} (y_i - y_j)^{2\beta} \prod\limits_{i} y_i^{u_+} (y_i - 1)^{v_+} \ f\big(y_1, \ldots, y_{N_-}\big)
\ee
and the constraint instead of (\ref{cnst}) is
\be\label{cnstb}
N_++N_-=1-\beta^{-1}-(2\beta)^{-1}(u_++u_-+v_++v_-)
\ee
We again use the transformation
\be\label{expb}
\prod\limits_{i = 1}^{N_+} (1 - w x_i)^{v_-} \prod\limits_{j = 1}^{N_-} (1 - w y_j)^{v_+} \prod\limits_{i = 1}^{N_+} \prod\limits_{j = 1}^{N_-} (1 - w x_i y_j)^{2\beta} = \emph{}\nn\\
\emph{} = \exp\left( - \beta \sum\limits_{k=1}^{\infty} \dfrac{w^k}{k} {\widetilde p}_k( p_k +\beta^{-1} v_+) \right) \exp\left( - \beta \sum\limits_{k=1}^{\infty} \dfrac{w^k}{k} p_k( {\widetilde p}_k + \beta^{-1}v_-) \right)
\ee
where $p_k = \sum_i x_i^k$ and ${\widetilde p}_k = \sum_i y_i^k$.

Now there are {\it two different} natural Cauchy expansions of this exponential into symmetric functions, which we discuss in the next two subsections.

\subsection{Double correlators of Jack polynomials}

First of all, one can naturally expand (\ref{expb}) into the Jack polynomials
\be\label{Jb}
 \exp\left( - \beta \sum\limits_{k=1}^{\infty} \dfrac{w^k}{k} {\widetilde p}_k( p_k +\beta^{-1} v_+) \right) \exp\left( - \beta \sum\limits_{k=1}^{\infty} \dfrac{w^k}{k} p_k( {\widetilde p}_k + \beta^{-1}v_-) \right)=\nn\\
=\sum\limits_{R',R''} w^{|R'|+|R''|} {J_{R'}\{ - p_k - \beta^{-1}v_+ \}J_{R''}\{p_k\}
J_{R'}\{ {\widetilde p}_k \} J_{R''}\{ -  {\widetilde p}_k - \beta^{-1}v_- \}\over ||R'||\cdot ||R''||}
\ee
where
\be
G_{R'R''}^\beta(x):=\prod_{(i,j)\in R'}\Big(x+R'_i-j+\beta(R''_j- i+1)\Big)
\ee
and
\be
||R||:={\overline{G}^\beta_{R^\vee R}(0)\over G^\beta_{RR^\vee}(0)}\beta^{|R|}
\ee
with the bar over the functions denoting the substitution $\beta\to\beta^{-1}$.

The quantity $G_{RR^\vee}^\beta(0)$ can be interpreted as a $\beta$-deformation of the product of the hook lengths over the Young diagram $R$, when length of the vertical part of the hook is multiplied by $\beta$. At $\beta=1$, $G^\beta_{RR^\vee}(0)$ is given by the usual hook formula, which is evident already from the identity $G_{RR^\vee}^\beta(0)=d_R^{-1}$.

The Jack polynomials, indeed, form a proper superintegrable basis for the Selberg $\beta$-ensemble, since the Selberg average of one Jack polynomial is factorized \cite{Kan,Kad}:
\be\label{sJc}
\Big<J_R\{p_k\}\Big> \ = \prod_{i=1}^N \int_0^1 dx_i x_i^u (1-x_i)^v \Delta^{2\beta}(x)J_R\Big\{p_k=\sum_ix_i^k\Big\}=
{J_R\{N\}J_R\{\beta^{-1}(u+1)+N-1\}\over
J_R\{\beta^{-1}(u+v+2)+2N-2\}}
\ee
Similarly, there is a factorization of double correlators of the Jack polynomials \cite{Kad2,MMShproof} (see also \cite{AFLT}):
\be
\Big<J_{R'}\{p_k+\rho\} J_{R''}\{p_k\}\Big>
= {\eta^\beta_{R'R''}\over \eta^\beta_{\emptyset \emptyset}}{J_{R'}\{v\beta^{-1}+N+\beta^{-1}-1\}
J_{R''}\{u\beta^{-1}+N+\beta^{-1}-1\}\over J_{R'}\{N\}J_{R''}\{u\beta^{-1}+N+\beta^{-1}-1+\rho\}}
\label{aveSpair}
\ee
where $\rho=\beta^{-1}v+\beta^{-1}-1$, and
\be
\eta^\beta_{R'R''}:={\prod_{i<j}^N(R'_i-i\beta-R'_j+j\beta)_\beta\cdot \prod_{i<j}^N(R''_i-i\beta-R''_j+j\beta)_\beta
\over \prod_{i,j}^N(u+v+2N\beta+2-\beta+R'_i-i\beta+R''_j-j\beta)_\beta}
\ee
with
\be
(x)_\beta:={\Gamma (x+\beta)\over\Gamma (x)}
\ee
being the Pochhammer symbol.

This expression can be also rewritten in the form
\be\label{dcJ}
\boxed{
\Big<J_{R''}\{p_k\} J_{R'}\{p_k+\rho\} \Big>
= J_{R'}\{N+\rho\}\Big<J_{R''}\{p_k\}\Big>
\times
\prod_{i=1}^N{J_{R'}\{\beta^{-1}(u+1+R''_i)+\rho+2N-1-i\}\over J_{R'}\{\beta^{-1}(u+1+R''_i)+\rho+2N-i\}}
}
\ee
Hence, the factorized average of properly normalized Jack polynomials is expressed completely in terms of the
same Jack polynomials, exactly in the spirit of superintegrability approach.

Note that this formula is not suitable for calculating the averages of products of Jack polynomials in (\ref{Jb}): first of all, the shift of $p_k$ in the Jack polynomials in (\ref{dcJ}) is different from that in (\ref{Jb}). Second,
an additional minus sign in the argument of $J_R$ in (\ref{Jb}) can not be remove just by transposition of the Young diagram as in the Schur case, since the transposition rule for the ordinary Jack polynomial is
\be\label{Jtr}
J_R\{-p_k\}=(-1)^{|R|}||R||\cdot\overline{J}_{R^\vee}\{\beta p_k\}
\ee
Because of it, expansion (\ref{Jb}) does not lead to factorized Nekrasov functions, while (\ref{dcJ}) does, but they do not correspond to the matrix model $Z_N^\beta(\{\alpha_a\},w)$, (\ref{3lb}).
As we shall see in the sext subsection, in order to deal with exponentials in (\ref{Jb}), one needs other, more involved bilinear combinations Jack polynomials, which still give rise to factorized expressions.

\subsection{Another deformation: generalized Jack polynomials}

Besides the $\beta$-deformation that promotes the Schur functions to the Jack polynomials, there is another one
that deforms just the whole product $S_{R''}\{p_k\}\cdot S_{R'}\{-p_k-v\}$ replacing the product with generalized Jack polynomials (GJP). Since the GJP are bilinear combinations of the Jack polynomials, one can just look at them as a proper candidate for bilinears that have factorized averages.

The GJP, $J_{R,P}\{p_k,\widetilde p_k|\lambda\}$ depends on two Young diagrams $R$ and $P$, on two sets of variables $p_k$ and $\widetilde p_k$, and on some spectral parameter $\lambda$. They are defined to be eigenfunctions of the Hamiltonian

\be
\hat H\, J_{R,P}\{p_k,\widetilde p_k|\lambda\}=\Lambda_{RP}(\lambda)\cdot J_{R,P}\{p_k,\widetilde p_k|\lambda\}
\ee
\be
\hat H=\beta \sum_{n,m}\left((n+m)p_np_m{\partial\over\partial p_{n+m}}
+(n+m)\widetilde p_n\widetilde p_m{\partial\over\partial \widetilde p_{n+m}}\right)
+\sum_{n,m}\left(nmp_{n+m}{\partial^2\over\partial p_n\partial p_m}
+nm\widetilde p_{n+m}{\partial^2\over\partial \widetilde p_n\partial \widetilde p_m}\right)+\nn\\
+(1-\beta)\sum_n\left((n-1)np_n{\partial\over\partial p_n}+(n-1)n\widetilde p_n{\partial\over\partial \widetilde p_n}\right)
+\lambda_1\sum_nnp_n{\partial\over\partial p_n}+\lambda_2\sum_nn\widetilde p_n{\partial\over\partial \widetilde p_n}
+2(1-\beta)\sum_nn^2p_n{\partial\over\partial\widetilde p_n}
\nn
\ee
with the eigenvalues
\be\label{ev}
\Lambda_{RP}(\lambda)=\lambda_1 |R|+\lambda_2 |P|+2\sum_{(i,j)\in R}\Big(j-1-\beta(i-1)\Big)+2\sum_{(i,j)\in P}\Big(j-1-\beta(i-1)\Big)
\ee
and $\lambda_1-\lambda_2=\lambda$.

This Hamiltonian is not self-adjoint because of the last term. In fact, it is clear that the set of conjugated functions is obtained from the GJP's by interchanging $(p_k,\lambda_1,R)\leftrightarrow (\widetilde p_k,\lambda_2,P)$ ($R$ and $P$ have to be interchanged because of formula (\ref{ev})), which means that the dual (adjoint) set of polynomials $J^*_{R,P}\{p_k,\widetilde p_k|\lambda\}$ is\footnote{This very simple relation is uplifted to a far more involved formula for the dual generalized Macdonald polynomials
\be
M^*_{R,P}\{p_k,\widetilde p_k|Q\}=
\left.M_{P,R}\left\{\widetilde p_k,p_k-\left(1-{t^{2n}\over q^{2n}}\right)\widetilde p_k\right|Q^{-1}\right\}
\nn
\ee
which follows from the Hamiltonian for the generalized Macdonald polynomials \cite{Zen,MMGen}
\be
\hat {\cal H}\{p,\bar p\} ={\rm res}_{z=0}\left\{
\exp\left(\sum_n \frac{(1-t^{-2n})z^n}{n}p_n\right)
\exp\left(\sum_{k>0}{q^{2k}-1\over z^k}
{\partial\over\partial p_k}\right)
+ \right. \nn \\ \!\!\!\!\!\!\!\!\!\!\!\!\! \left.
+ Q^{-1}\cdot \exp\left(\sum_n \frac{(1-t^{-2n})z^n}{n}
\left[\left(1-{t^{2n}\over q^{2n}}\right)p_n+\widetilde p_n\right]\right)
\exp\left(\sum_{k>0}{q^{2k}-1\over z^k}
{\partial\over\partial \widetilde p_k}\right)
\right\} \ \ \ \ \
\label{GEMham}\nn
\ee
}
\be\label{dual}
J^*_{R,P}\{p_k,\widetilde p_k|\lambda\}:=J_{P,R}\{\widetilde p_k,p_k|-\lambda\}
\ee
These dual GJP's are orthogonal to the GJP's,
\be
\Big<J_{R,P}\{p_k,\widetilde p_k|\lambda\},J^*_{R',P'}\{p_k,\widetilde p_k|\lambda\}\Big>=||R||\cdot ||P||\cdot\delta_{RR'}\delta_{PP'}
\ee
w.r.t. the scalar product
\be
\Big< p_{\Delta_1}\widetilde p_{\Delta_2}\Big| p_{\Delta'_1}\widetilde p_{\Delta'_2}\Big>=\beta^{-l_{\Delta_1}-l_{\Delta_1}}
z_{\Delta_1}z_{\Delta_2}\delta_{\Delta_1\Delta_1'}\delta_{\Delta_2\Delta_2'}
\ee
Thus, we normalize the GJP in such a way that the Cauchy identity looks like
\be\label{CauchyGJP}
\sum _{R,P}\ {J_{R,P}\{p_k,\widetilde p_k|\lambda\}\cdot J^*_{R,P}\{p'_k,\widetilde p'_k|\lambda\}\over ||R||\cdot ||P||}=
\sum _{R,P}\ {J_{R,P}\{p_k,\widetilde p_k|\lambda\}\cdot J_{P,R}\{\widetilde p'_k, p'_k|-\lambda\}\over ||R||\cdot ||P||}=
\exp\left(\beta \sum_k{p_kp'_k+\widetilde p_k\widetilde p'_k\over k}\right)
\ee
The GJP's turn into the product of two Schur functions at $\beta=1$:
\be
J_{R,P}\{p_k,\widetilde p_k|\lambda\}\Big|_{\beta=1}=S_R\{p_k\}\cdot S_P\{\widetilde p_k\}
\ee
so that the dependence on the spectral parameter disappears. Hence, the GJP can play a role of a deformation of the product of two Schur functions. Now one can check that the superintegrability also survives for the GJP's. In this case, one evaluates the matrix model (Selberg) average of one GJP, which is a deformation of the product of two Schur functions.
The GJP counterpart ($\beta$-deformation) of formula (\ref{dcor3}) becomes in this case (see also \cite{MS})
\be\label{dGJ+}
\!\!
\boxed{
\begin{array}{c}
\Big<J_{R',R''}\{p_k,-p_k-\beta^{-1}v|\lambda_*\}\Big>=\\
\cr
\!\!\!=\!
\displaystyle{{(-1)^{|R'|+|R''|}||R'||\cdot ||R''||\over \beta^{2|R''|}}
{J_{R'}\{\beta^{-1}(u+1)+N-1\}J_{R'}\{N\} \overline{J}_{R''^\vee}\{N\beta+v\}\overline{J}_{R''^\vee}\{\beta (N-1)+u+v+1\}\over {\cal F}^\beta_{R''^\vee R'}(\beta^{-1}(u+v+1)+2N-1)J_{R'}^2\{\delta_{k,1}\}\overline{J}_{R''^\vee}^2\{\delta_{k,1}\}}}
\!\!\!\!
\end{array}}
\ee
where
\be
\lambda_*:=4N\beta+2(u+v+1-\beta)
\ee
and
\be
{\cal F}^\beta_{R''R'}(x):=
\overline{G}^\beta_{R''R''^\vee}(0)\overline{G}^\beta_{R''R'}(x)
\overline{G}^\beta_{R'^\vee R''^\vee}(1-\beta-x)\overline{G}^\beta_{R'^\vee R'}(0)
\ee
Similarly to (\ref{dcJ}), this formula is also expressed completely in terms of the
Jack polynomials in the spirit of superintegrability approach:
\be\label{dGJ+J}
\!\!\!\!
\boxed{
\Big<J_{R'',R'}\{p_k,-p_k-\bar\rho|\lambda_*\}\Big>
= J_{R'}\{-N-\bar\rho\}\Big<J_{R''}\{p_k\}\Big>
\times
\prod_{i=1}^N{J_{R'}\{-\beta^{-1}(u+1+R''_i)-\bar\rho-2N+1+i\}\over J_{R'}\{-\beta^{-1}(u+1+R''_i)-\bar\rho-2N+i\}}
}
\ee
where we denoted $\bar\rho:=\beta^{-1}v$ in order to emphasize a resemblance of this formula and (\ref{dcJ}).
Indeed, this expression is just (\ref{dcJ}) with signs of arguments of $J_{R'}$ properly changed and with the replace $\rho\to\bar\rho$.

In the case of GJP, the transposition rule requires changing signs of the {\it both} sets of times:
\be
J_{R^\vee,P^\vee}\{p_k,\widetilde p_k|\lambda\}=
(-1)^{l_R+l_P}\cdot||R||\cdot ||P||\cdot\overline{J}_{R,P}\{-\beta p_k,-\beta \widetilde p_k|-\beta^{-1}\lambda\}
\ee
and the average $\Big<J_{R',R''}\{p_k,p_k+const|\lambda\}\Big>$ is not immediate to factorize.

Similarly, as follows from (\ref{dual}),
\be\label{dGJ-}
\boxed{
\Big<J^*_{R',R''}\{-p_k-\beta^{-1}v,p_k|-\lambda_*\}\Big>=\Big<J_{R'',R'}\{p_k,-p_k-\beta^{-1}v|\lambda_*\}\Big>}
\ee

In the Appendix, we explain how one can technically check in concrete examples that these correlators are indeed factorized.

\subsection{Nekrasov functions and AGT}

Now we are ready to present the three-logarithm $\beta$-ensemble as a sum over factorized terms using the GJP. Indeed,
\be
Z_{N_1,N_2}^\beta(u_\pm,v_\pm;w)\stackrel{(\ref{expb})}{=}
\left< \left< \exp\left( - \beta \sum\limits_{k=1}^{\infty} \dfrac{w^k}{k} {\widetilde p}_k( p_k +\beta^{-1} v_+) \right) \exp\left( - \beta \sum\limits_{k=1}^{\infty} \dfrac{w^k}{k} p_k( {\widetilde p}_k + \beta^{-1}v_-) \right) \right>_+ \right>_-
\stackrel{(\ref{CauchyGJP})}{=}\nn\\
=\sum_{R',R''}{w^{|R'|+|R''|}\over ||R'||\cdot ||R''||}
\cdot\Big<J_{R',R''}\{p_k,-p_k-\beta^{-1}v|\lambda_+\}\Big>_+
\cdot\Big<J^*_{R',R''}\{-p_k-\beta^{-1}v,p_k|\lambda_+\}\Big>_-
\ \ \ \ \ \ \ \ \ \ \
\ee
where we choose the spectral parameter (which can be freely chosen) to be $\lambda_+:=4N\beta+2(u_++v_++1-\beta)$. Note that it follows from (\ref{cnstb}) that this $\lambda_+=
-4N\beta-2(u_-+v_-+1-\beta)=-\lambda_-$. Hence, one can finally write
\be\label{zib}
Z_{N_1,N_2}^\beta(u_\pm,v_\pm;w)=\sum_{R',R''}w^{|R'|+|R''|}\overbrace{{\left<J_{R',R''}\{p_k,-p_k-\beta^{-1}v|\lambda_+\}\right>_+
\cdot\left<J^*_{R',R''}\{-p_k-\beta^{-1}v,p_k|-\lambda_-\}\right>_-\over ||R'||\cdot ||R''||}}^{\rm Nekrasov\ functions}
\stackrel{(\ref{dGJ+}),(\ref{dGJ-})}{=}\ \ \nn\\
\!\!\!\!\!\!\!\!\!\!\!\!\!
=\sum_{R',R''}{(\beta^{-2}w)^{|R'|+|R''|}\over ||R'||^{-1}||R''||^{-1}}
\cdot {J_{R'}\{\beta^{-1}(u_++1)+N_+-1\}J_{R'}\{N_+\} \overline{J}_{R''^\vee}\{N_+\beta+v_+\}\overline{J}_{R''^\vee}\{\beta (N_+-1)+u_++v_++1\}\over {\cal F}^\beta_{R''^\vee R'}(\beta^{-1}(u_++v_++1)+2N_+-1)J_{R'}^2\{\delta_{k,1}\}\overline{J}_{R''^\vee}^2\{\delta_{k,1}\}}
\times\ \nn
\ee
\vspace{-0.3cm}
\be
\times
{J_{R''}\{\beta^{-1}(u_-+1)+N_--1\}J_{R''}\{N_-\} \overline{J}_{R'^\vee}\{N_-\beta+v_-\}\overline{J}_{R'^\vee}\{\beta (N_--1)+u_-+v_-+1\}\over {\cal F}^\beta_{R'^\vee R''}(\beta^{-1}(u_-+v_-+1)+2N_--1)J_{R''}^2\{\delta_{k,1}\}\overline{J}_{R'^\vee}^2\{\delta_{k,1}\}}
\ee
Using the notations
\be
2\mu_1:={v_+-u_+-1\over\beta}+1\ \ \ \ \
2\mu_2:={v_++u_++1\over\beta}-1\ \ \ \ \
2\mu_3:={v_--u_--1\over\beta}+1\nn\\
2\mu_4:={v_-+u_-+1\over\beta}-1\ \ \ \ \
\mu:=N_++\mu_2=-N_--\mu_4
\ee
and
\be
z_f^\beta(R,x):={\overline{J}_R\{p_k=\beta x\}\over \overline{J}_R\{\delta_{k,1}\}}
\ee
one rewrites (\ref{zib}) in the standard Nekrasov form \cite{MMShproof}

\be\label{LMNS}
Z_{N_1,N_2}^\beta(u_\pm,v_\pm;w)=\sum_{R',R''}{(\beta^{-2}w)^{|R'|+|R''|}\over ||R'||^{-1}||R''||^{-1}}
\cdot{J_{R'}\{\mu-\mu_1\}J_{R'}\{\mu-\mu_2\} \overline{J}_{R''^\vee}\{\beta(\mu+\mu_1)\}\overline{J}_{R''^\vee}\{\beta (\mu+\mu_2)\}\over {\cal F}^\beta_{R''^\vee R'}(2\mu)J_{R'}^2\{\delta_{k,1}\}\overline{J}_{R''^\vee}^2\{\delta_{k,1}\}}
\times\nn
\ee
\vspace{-0.3cm}
\be
\times
{J_{R''}\{-\mu-\mu_3\}J_{R''}\{-\mu-\mu_4\} \overline{J}_{R'^\vee}\{-\beta(\mu-\mu_3)\}\overline{J}_{R'^\vee}\{-\beta (\mu-\mu_4)\}\over {\cal F}^\beta_{R'^\vee R''}(-2\mu)J_{R''}^2\{\delta_{k,1}\}\overline{J}_{R'^\vee}^2\{\delta_{k,1}\}}
\stackrel{(\ref{Jtr})}{=}\nn\\
=\sum_{R',R''}{(\beta^{-2}w)^{|R'|+|R''|}\over ||R'||^{-3}||R''||^{-3}}
\cdot{\prod_{a=1}^4z_f^\beta(R''^\vee,\mu_a+\mu)z_f^\beta(R'^\vee,\mu_a-\mu)\over {\cal F}^\beta_{R''^\vee R'}(2\mu) {\cal F}^\beta_{R'^\vee R''}(-2\mu)}\cdot{\overline{J}_{R',^\vee}^2\overline{J}_{R''^\vee}^2\over J_{R'}^2\{\delta_{k,1}\}J_{R''}^2\{\delta_{k,1}\}}=\nn\\
=\sum_{R',R''}{(\beta^{-4}w)^{|R'|+|R''|}\over ||R'||^{-1}||R''||^{-1}}
\cdot{\prod_{a=1}^4z_f^\beta(R''^\vee,\mu_a+\mu)z_f^\beta(R'^\vee,\mu_a-\mu)\over
 {\cal F}^\beta_{R''^\vee R'}(2\mu) {\cal F}^\beta_{R'^\vee R''}(-2\mu)}=
 \nn\\
 = \sum_{R',R''}w^{|R'|+|R''|}
\cdot{\prod_{a=1}^4z_f^\beta(R'',\mu_a+\mu)z_f^\beta(R',\mu_a-\mu)\over z_{vect}^\beta(R'',R'^\vee,2\mu)}
\ee
since
\be
J_{R}\{\delta_{k,1}\}=\beta^{|R|}||R||\cdot \overline{J}_{R^\vee}\{\delta_{k,1}\}
\ee
and
\be
z_{vect}^\beta(R'',R'^\vee,2\mu)={\beta^{4|R'|+4|R''|}\over ||R''^\vee||\cdot ||R'||}
{\cal F}^\beta_{R'' R'}(2\mu) {\cal F}^\beta_{R'^\vee R''^\vee}(-2\mu)
\ee

\section{More examples of DV models\label{newm}}

\noindent
Now we will apply our general construction
\be
%\boxed{
{\bf
factorized\ correlator\ of\ character\ bilinears   \longrightarrow   DV\ block
 \longrightarrow   Nekrasov\ function \longrightarrow   matrix\ model
 }
% }
\nn
\ee
to other factorized correlators in order to construct  new
solvable eigenvalue matrix models.

\subsection{Models of logarithm type}

First of all, using (\ref{dcJ}), one can easily deal with the models (one needs carefully deals with the poles in this expression, see, e.g., \cite{MMM}):
\be\label{Z1}
Z_{N_1,N_2}^{(1)}(u_\pm,v_\pm)=\prod_{i=1}^{N_+} \int dx_i x_i^{u_-} (1-x_i)^{v_-}(1-wx_i)^{-\rho_-} \Delta^{2\beta}(x)
\times \nn\\
\times \prod_{j=1}^{N_-} \int dy_j y_j^{u_+} (1-y_j)^{v_+}(1-wy_j)^{-\rho_+} \Delta^{2\beta}(y)
\times \prod_{i,j}(1-wx_iy_j)^{-2\beta}=
\nn\\
=\left< \ \left< \ \  \exp\left(  \beta \sum\limits_{k=1}^{\infty} \dfrac{w^k}{k} {\widetilde p}_k( p_k +\rho_+) \right) \exp\left( \beta \sum\limits_{k=1}^{\infty} \dfrac{w^k}{k} p_k( {\widetilde p}_k + \rho_-) \right)\ \right>_+ \ \right>_-=\nn\\
=\sum\limits_{R',R''} w^{|R'|+|R''|} \underbrace{{\Big<J_{R'}\{ p_k +\rho_+ \}J_{R''}\{p_k\}\Big>_+
\Big<J_{R''}\{ {\widetilde p}_k +\rho_- \}J_{R'}\{ {\widetilde p}_k \}\Big>_-\over ||R'||\cdot ||R''||}}_{\rm Nekrasov\ functions}
\ee
and
\be\label{Z2}
Z_{N_1,N_2}^{(2)}(u_\pm,v_\pm;\xi_k,\eta_k)=(1-w)^{-\beta \rho_+\rho_-}
\prod_{i=1}^{N_+} \int dx_i x_i^{u_-} (1-x_i)^{v_-}(1-wx_i)^{-\rho_-} \Delta^{2\beta}(x)\exp\left( \beta \sum_{k=1,i}x_i^k\xi_k \right)
\times \nn
\ee
\vspace{-0.7cm}
\be
\times \prod_{j=1}^{N_-} \int dy_j y_j^{u_+} (1-y_j)^{v_+}(1-wy_j)^{-\rho_+} \Delta^{2\beta}(y)
\exp\left( \beta \sum_{k=1,i}y_i^k\eta_k \right)\times \prod_{i,j}(1-wx_iy_j)^{-\beta}=
\nn\\
=\left< \ \left< \ \  \exp\left(  \beta \sum\limits_{k=1}^{\infty} \dfrac{w^k}{k} ({\widetilde p}_k+\rho_-)( p_k +\rho_+) \right) \exp\left( \beta \sum\limits_{k=1}^{\infty} (p_k\xi_k+ {\widetilde p}_k \eta_k) \right)\ \right>_+ \ \right>_-=\nn\\
=\sum\limits_{R',R'',R'''} w^{|R'|} \underbrace{{\Big<J_{R'}\{ p_k +\rho_+ \}J_{R''}\{ {p}_k \}\Big>_+
\Big<J_{R'}\{ {\widetilde p}_k +\rho_- \}J_{R'''}\{\widetilde p_k\}\Big>_- J_{R''}\{ {\xi}_k \}J_{R'''}\{{\eta}_k \}
\over ||R'||\cdot ||R''||\cdot ||R'''||}}_{\rm Nekrasov\ functions}
\ee
where $\rho_\pm:=+\beta^{-1}(v_\pm+1)-1$, and $\xi_k$, $\eta_k$ are arbitrary constants. Note that the triple summation arises here since the partition function includes the sources $\xi_k$, $\eta_k$, which generate all invariant correlation functions. The partition function without the sources reduces to just a single sum
\be
Z_{N_1,N_2}^{(2)}(u_\pm,v_\pm;0,0)=(1-w)^{-\beta \rho_+\rho_-}
\prod_{i=1}^{N_+} \int dx_i x_i^{u_-} (1-x_i)^{v_-}(1-wx_i)^{-\rho_-} \Delta^{2\beta}(x)\times \prod_{i,j}(1-wx_iy_j)^{-\beta}
\times \nn\\
\times \prod_{j=1}^{N_-} \int dy_j y_j^{u_+} (1-y_j)^{v_+}(1-wy_j)^{-\rho_+} \Delta^{2\beta}(y)
=\sum\limits_{R'} w^{|R'|} \underbrace{{\Big<J_{R'}\{ p_k +\rho_+ \}\Big>_+
\Big<J_{R'}\{ {\widetilde p}_k +\rho_- \}\Big>_-
\over ||R'||}}_{\rm Nekrasov\ functions}\ \ \ \ \ 
\ee

These are eigenvalue models, and the second model is the model of the type considered earlier within the framework of conformal matrix models, whose partition functions satisfy $W$-algebra constraints \cite{MMM,confmamo} (see also \cite{Kostov}), and they are described by the Nekrasov functions following from (\ref{dcJ}).

Another interesting set of models arise when one changes
the mixing factor $1-x_iy_j$ in above models for $x_i-y_j$
(again, one has to be careful with the poles).
This
gives rise to
$p_{-k}=\sum_ix_i^{-k}$ and $\widetilde p_{-k}=\sum_iy_i^{-k}$ in the exponentials (\ref{Z1}), (\ref{Z2}),
and the corresponding Nekrasov functions are based on the factorized Selberg correlators related to (\ref{dcJ}),
\be
\boxed{
\Big<J_{R'}\{p_k+\rho\}J_{R''}\{p_{-k}\}\Big>=\Big<J_{R'}\{p_k+\rho\}J_{\widetilde R''}\{p_{k}\}\Big>\Big|_{u\to u-R''_1}
}
\ee
after using the identity by Kadell \cite{Kad}:
\be
J_R\{p_{-k}\}=\prod_{i=1}^Nx_i^{-R_1}\cdot J_{\widetilde R}\{p_k\}
\ee
where $p_{-k}=\sum_i^Nx_i^{-k}$ and $\widetilde R_i:=R_1-R_{N-i+1}$.

\subsection{Model with Gaussian DV building blocks\label{Gauss}}

All the models we considered so far were based on DV blocks with the Selberg measure. Now we consider the case of Gaussian measure, where the bilinears in Schur functions are more involved, and the whole construction is looking more complicated. Still, it exists and can be dealt with.

The starting point is constructing bilinears \cite{MMdsi}, which are the products
\be
K_\Delta\{\Tr H^k\}\cdot S_R\{\Tr H^k\},\ \ \ \ \ \ \ \hbox{with}\ \ \
K_\Delta\{\Tr H^k\}=(-1)^{|\Delta|}\ e^{\frac{1}{2}\tr H^2}\, \hat W^-_\Delta\, e^{-\frac{1}{2}\tr H^2}
\ee
where\footnote{See also \cite{Ch} on another realization of these operators, which better suits the framework of \cite{MMMR}.}
\be
\hat W^-_k:=\Tr\left( \frac{\p}{\p H}\right)^k\nn\\
\hat W^-_\Delta:=\prod_a^{l_\Delta}W^-_{\Delta_a}
\ee
and $H$ is an $N\times N$ Hermitian matrix.
The quantities $K_\Delta\{\Tr H^k\}$ are linear combinations of the Schur functions that form a full basis in the space of polynomials of $\Tr H^k$. The crucial point is that the averages of $K_\Delta\{\Tr H^k\}\cdot S_R\{\Tr H^k\}$ are factorized with the Gaussian measure:
\be
\Big<\ldots\Big>_{GH}:=\int dH \ldots\exp\left(-{1\over 2}\Tr H^2\right),\ \ \ \ \ \
\Big<1\Big>:=1\nn\\
\Big<K_\Delta\{\Tr H^k\}\cdot S_R\{\Tr H^k\}\Big>_{GH}=\mu_{R,\Delta}\Big<S_R\{\Tr H^k\}\Big>_{GH}=
\mu_{R,\Delta}{S_R\{N\}S_R\{\delta_{k,2}\}\over S_R\{\delta_{k,1}\}}
\ee
where $\mu_{R,\Delta}$ are numbers that do not depend on the size of matrix $N$, see \cite{MMdsi} for details. In this formula, $P$ is even, since otherwise both the average of $S_R$ vanishes and $\mu_{R,\Delta}$ becomes singular. However, after rewriting it in another form, see formula (\ref{fK}) below, this restriction is lifted.

In fact, one can naturally consider the linear combinations of $K_\Delta$:
\be
\chi_P:=\sum_{\Delta}{\psi_P(\Delta)K_\Delta\over z_\Delta}=
(-1)^{|P|}\ e^{\frac{1}{2}\tr H^2}\, S_P\{\hat W^-_k\}\, e^{-\frac{1}{2}\tr H^2}
\ee
where $\psi_P(\Delta)$ is the character of symmetric group ${\cal S}_{|P|}$ in representation $P$, and $z_\Delta$ is the standard symmetric factor of the Young diagram (order of the automorphism).
If $m_k$ is the number of lines of length $k$ in the Young diagram $\Delta$, then $z_\Delta:=\prod_k k^{m_k}m_k!$.

The quantities $\chi_P$ are generated by the operators $S_P\{\hat W^-_k\}$, and the averages now take the form
\be\label{fK}
\boxed{
\begin{array}{c}
\Big<S_P\{\hat W^-_k\}\ S_R\{\Tr H^k\}\Big>_{GH}=\int dH \exp\left(-{1\over 2}\Tr H^2\right)
S_P\{\hat W^-_k\}\ S_R\{\Tr H^k\}=\\
\\
=\Big<\chi_P\{\Tr H^k\}\cdot S_R\{\Tr H^k\}\Big>_{GH}={\displaystyle{S_{R/P}\{\delta_{k,2}\}}\over \displaystyle{S_{R}\{\delta_{k,2}\}}}\ \Big<S_R\{\Tr H^k\}\Big>_{GH}=
{\displaystyle{S_{R/P}\{\delta_{k,2}\}S_R\{N\}}\over \displaystyle{S_R\{\delta_{k,1}\}}}
\end{array}}
\ee
where we integrated by parts in the second equality. This average certainly vanishes when $|R|-|P|$ is odd (but $|R|$ and $|P|$ can be both odd giving rise to a non-zero contribution).

Hence, these averages give us DV blocks, and one can use them to construct the Nekrasov functions and then a matrix model:
\be
Z^G_{N_1,N_2}:=\sum_{R',R''}w^{|R'|+|R''|}\underbrace{\Big<\chi_{R'}\{\Tr X^k\}\cdot S_{R''}\{\Tr X^k\}\Big>
\Big<\chi_{R'}\{\Tr Y^k\}\cdot S_{R''}\{\Tr Y^k\}\Big>}_{\rm Nekrasov\ functions}
\ee
It is a two-matrix integral over $N_+\times N_+$ Hermitian matrix $X$ and $N_-\times N_-$ Hermitian matrix $Y$:
\be
Z^G_{N_1,N_2}= \int DXDY e^{-\frac{1}{2}\tr X^2}e^{-\frac{1}{2}\tr Y^2}
\sum_{R''} w^{|R''|} S_{R''}\{\tr X^k\}S_{R''}\{\tr Y^k\} \sum_{R'} w^{|R'|}
\chi_{R'}\{\tr X^k\}\chi_{R'}\{\tr Y^k\}
= \nn \\
= \int DXDY \exp\left(\sum_k{\Tr X^k\,\Tr Y^k\over k}w^k\right)\sum_{R'} w^{|R'|}S_{R'}\{\hat W^-_k(X)\}S_{R'}\{\hat W^-_k(Y)\}
 e^{-\frac{1}{2}\tr X^2}  e^{-\frac{1}{2}\tr Y^2}=\nn\\
 = \int DXDY \exp\left(\sum_k{\Tr X^k\,\Tr Y^k\over k}w^k\right)\exp\left(\sum_k {w^k\over k}\Tr\left( \frac{\p}{\p X}\right)^k
\Tr\left( \frac{\p}{\p Y}\right)^k\right) e^{-\frac{1}{2}\tr X^2}  e^{-\frac{1}{2}\tr Y^2}
\ee

\section{Conclusion}

In this paper, we explained that {\bf the very existence of Nekrasov functions},
i.e. factorization of the average of point-split {\it pair} correlators is a peculiar property of superintegrability.
We introduced a notion of the DV block, which is defined by the input pair: ``measure'' and ``bilinear combination of characters'' such that averaging of this bilinear combination with this measure is factorized. This allowed us {\bf to extend the notion of Nekrasov function and ascribe this name to the product of two arbitrary DV blocks}. Thus defined Nekrasov functions give rise to a matrix or eigenvalue model, on one hand, and the class of Nekrasov functions becomes well defined and well structured, on the other hand.

Thus, our construction is:
\be
\boxed{
\text{factorized correlator of character bilinears }  \longrightarrow  \text{ DV block}
 \longrightarrow  \text{ Nekrasov function} \longrightarrow  \text{ matrix model}
 }
\nn
\ee
In section 5, we constructed in such a way various new models that can be solved.

\bigskip

In this paper, we considered four examples of factorized averages of bilinears of characters:
\begin{itemize}
\item the Selberg average, when the bilinear is just $S_{R'}\{p_k\}\cdot S_{R''}\{p_k+v\}$;
\item the $\beta$-deformed Selberg average, when there are two factorized averages of
bilinears of Jack polynomials:
\begin{itemize}
\item[{\bf i)}]
a simple bilinear $J_{R'}\{p_k\}\cdot J_{R''}\{p_k+\beta^{-1}(v+1)-1\}$ and
\item[{\bf ii)}]
a more complicated linear combination
of Jack bilinears,
the generalized Jack polynomials;
\end{itemize}
\item the Gaussian average, when the bilinear is $\chi_P\{p_k\}\cdot S_R\{p_k\}$, and $\chi_P$ is a peculiar linear combination of the Schur functions.
\end{itemize}
In all these cases we observe that the factorized average of bilinears is expressed through the properly normalized characters \cite{Mac}: see formulas (\ref{dcJ}), (\ref{dGJ+J}) and (\ref{fK}). Hence, we conjecture that this is a general phenomenon: {\bf factorized averages of bilinears of properly normalized characters are expressed through the same characters}.

Our approach can be embedded into an algebraic framework. Indeed, the Nekrasov functions we described in this paper are associated with the $SU(2)$ group. In the case of arbitrary $SU(N)$ group, the Nekrasov function is still constructed as a product of two DV blocks, however, the DV block is now not a bilinear combination of characters, but an $N$-linear one. More precisely, one has to introduce $N-1$ different measures $\mu_a$, $a=1,\ldots,N-1$, each one associated with a set of variables $\{p_k^{(a)}\}$, and consider a linear combination of products of the form $\prod_{a=0}^{N-1}\chi_a\{p_k^{(a+1)}-p_k^{(a)}\}$, $p_k^{(0)}=p_k^{(N)}=0$, where $\chi$ denotes proper characters (for instance, the Schur functions, or the Jack polynomials) \cite{MMZ}. These linear combinations are chosen in such a way that they are factorized after averaging with $\{\mu_a\}$, which leads us to a DV block. If one now at the matrix model that is generated by the constructed Nekrasov function, it becomes an essentially multi-matrix model.

Extensions to other root systems of Lie algebras, and to quivers are immediate. In particular, in order to deal with the affine algebra $\hat A_{N-1}$, it is enough to switch from
the Dirichlet boundary conditions $p_k^{(0)}=p_k^{(N)}=0$  to the periodic ones $p_k^{(0)}=p_k^{(N)}$.

\section*{Acknowledgements}

Our work is partly supported by the grant of the Foundation for the Advancement of Theoretical Physics ``BASIS'' (A.Mor.), by joint grants 21-52-52004-MNT-a and 21-51-46010-ST-a.

\section*{Appendix}

 A general reason and a proof of factorization in formula (\ref{dGJ+J}) are not present in this paper. Let us, however, explain how one can technically check that the averages (\ref{dcJ}) and (\ref{dGJ+J}) are indeed factorized in concrete examples.
 
First of all, in order to calculate $\Big<J_{R''}\{p_k\}J_{R'}\{p_k+\alpha\}\Big>$ with some parameter $\alpha$, one can expand the Jack polynomials into the skew Jack polynomials,
\be\label{shJ}
J_{R'}\{p_k+\alpha\}=\sum_Q J_{R'/Q}\{\alpha\}J_Q\{p_k\}
\ee
and use the expansion
\be
J_{R''}\{p_k\}J_Q\{p_k\}=\sum_{P}{\bf N}^P_{R''Q}J_P\{p_k\}
\ee
where ${\bf N}^P_{R''Q}$ are the $\beta$-dependent Littlewood-Richardson coefficients, in order to express the product $J_{R''}\{p_k\}J_{R'}\{p_k+\alpha\}$ through the single ordinary Jack polynomial, its average being given by (\ref{sJc}). This would be enough to calculate the averages of the products of two Jack polynomials as in (\ref{dcJ}).
 
The simplest way to evaluate the GJP average is to expand the GJP into products of pairs of the ordinary Jack polynomials at the two different sets of time variables,
\be\label{dJ}
J_{R',R''}\{p_k,\widetilde p_k|\lambda\}=\sum_{Q',Q''}c_{R'R''}^{Q'Q''}(u)J_{Q'}\{p_k\}J_{Q''}\{\widetilde p_k\}
\ee
Now there is an additional sign minus in front of $p_k$, and one has to re-expand $J_R\{-p_k\}$ to $J_R\{p_k\}$, using the orthogonality of the Jack polynomials
\be
\Big<J_{R'}\{p_k\}\Big|J_{R''}\{p_k\}\Big>=||R'||\cdot\delta_{R'R''}
\ee
w.r.t. to the scalar product
\be
\Big<p_\Delta\Big| p_{\Delta'}\Big>=\beta^{-l_\Delta}z_\Delta\delta_{\Delta,\Delta'}
\ee
hence obtaining
\be
J_R\{-p_k\}=\sum_Q ||Q||^{-1}\Big<J_{R}\{-p_k\}\Big|J_{Q}\{p_k\}\Big>\cdot J_Q\{p_k\}
\ee
Thus, finally one obtains
\be
J_{R'}\{-p_k-\beta^{-1}v\}J_{R''}\{p_k\}=\sum_{Q'} J_{R'/Q'}\{-\beta^{-1}v\}J_{Q'}\{-p_k\}J_{R''}\{p_k\}=\nn\\
=\sum_{Q',Q''} J_{R'/Q'}\{-\beta^{-1}v\}||Q''||^{-1}\Big<J_{Q'}\{-p_k\}\Big|J_{Q''}\{p_k\}\Big>\cdot J_{Q''}\{p_k\}J_{R''}\{p_k\}=
\nn\\ =\sum_{Q',Q''} J_{R'/Q'}\{-\beta^{-1}v\}||Q''||^{-1}\Big<J_{Q'}\{-p_k\}\Big|J_{Q''}\{p_k\}\Big>{\bf N}^P_{Q''R''}\cdot J_{P}\{p_k\}
\ee
and, inserting this formula into (\ref{dJ}) and evaluating further the average of the single Jack polynomial $J_{P}\{p_k\}$ using (\ref{dcJ}), one can check the factorization of the GJP average in concrete examples.

\end{document}